\documentclass[10pt,aps,prl,twocolumn,superscriptaddress]{revtex4-2}

\usepackage{xcolor}
\usepackage{lipsum}
\newsavebox{\figurebox}

\usepackage[T1]{fontenc}
\usepackage{amsmath}
\usepackage{graphicx}
\usepackage{dcolumn}
\usepackage{cmupint} 
\usepackage[explicit]{titlesec}

\usepackage[hidelinks,colorlinks=true,
            citecolor=blue, 
            linkcolor=blue, 
            urlcolor=blue 
            ]{hyperref}
\hypersetup{
    pdftitle = {gancewski2026},
    pdfauthor = {M. Gancewski, E. Adkins, H. Jóźwiak, N. Stolarczyk, E. Quintas-S\'{a}nchez, R. Dawes, J. T. Hodges, P. Wcisło},
}

\let\bs\boldsymbol

\titleformat{\paragraph}[runin]{\itshape}{}{0em}{#1---}

\begin{document}

\title{
    \texorpdfstring{Collisional lineshapes of an N$_2$-perturbed electronic transition in O$_2$ from first principles of quantum theory validated by accurate experimental spectra
    }{}
}

\author{Maciej Gancewski}
\affiliation{
    Institute of Physics, Faculty of Physics, Astronomy and Informatics, Nicolaus Copernicus University in Toru\'{n}, ul. Grudzi\k{a}dzka 5, 87-100 Toru\'{n}, Poland
}

\author{Erin Adkins}
\affiliation{
    Chemical Sciences Division, National Institute of Standards and Technology, Gaithersburg, MD 20899, U.S.A.
}

\author{Hubert J\'{o}\'{z}wiak}
\author{Nikodem Stolarczyk}
\affiliation{
    Institute of Physics, Faculty of Physics, Astronomy and Informatics, Nicolaus Copernicus University in Toru\'{n}, ul. Grudzi\k{a}dzka 5, 87-100 Toru\'{n}, Poland
}

\author{Ernesto Quintas-S\'{a}nchez}
\author{Richard Dawes}
\affiliation{
    Department of Chemistry, Missouri University of Science and Technology, Rolla, MO 65409-0010, USA
}

\author{Joseph T. Hodges}
\affiliation{
    Chemical Sciences Division, National Institute of Standards and Technology, Gaithersburg, MD 20899, U.S.A.
}

\author{Piotr Wcis{\l}o}
\affiliation{
    Institute of Physics, Faculty of Physics, Astronomy and Informatics, Nicolaus Copernicus University in Toru\'{n}, ul. Grudzi\k{a}dzka 5, 87-100 Toru\'{n}, Poland
}

\begin{abstract}
    We demonstrate the feasibility of predicting the lineshape of an electronic molecular transition collisionally perturbed by another molecule from first principles of quantum mechanics. We consider an N$_2$-perturbed O$_2$ \textit{A}-band transition, which is relevant to spectroscopic observations in Earth's atmosphere.  This approach simultaneously incorporates the electronic structure, quantum scattering and lineshape effects required to accurately predict pressure- and temperature-dependence of observed lineshapes. The \textit{ab initio} lineshapes reproduce measurements to better than 0.6\%. Theory  also provides new physical constraints on beyond-Voigt lineshape parameters in fits of high-precision data, yielding fit residuals at the 0.08\% level. This work motivates advances in molecular electronic structure calculations and demonstrates the potential to generate comprehensive spectroscopic reference data from first principles.
\end{abstract}

\maketitle

\begin{figure}
    \centering
    \includegraphics[width=\linewidth]{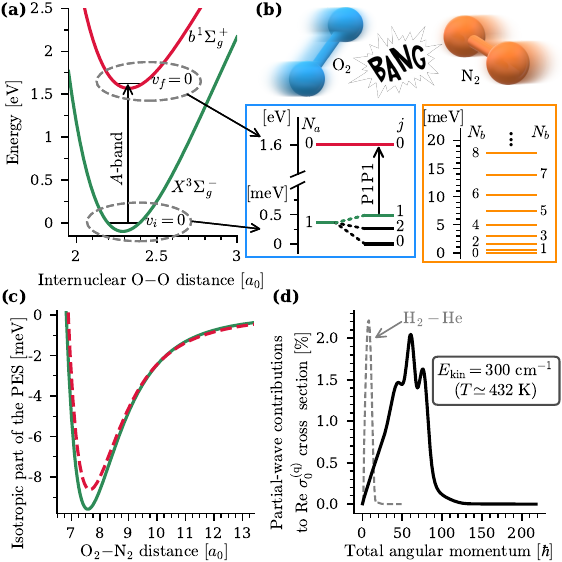}
    \caption{
        \label{fig: fig 1}
        (a)~The O$_2$ \textit{A}-band comprises rovibronic transitions between the lowest indicated vibrational $v_i=v_f=0$ levels of the electronic terms $X^3\Sigma^-_g$ (green curve) and $b^1\Sigma^+_g$ (red curve) of O$_2$. (b)~(left blue box): P1P1 \textit{A}-band transition (arrow) connecting the fine structure-resolved $|X^3\Sigma^-_g, N_a=1, j=1\rangle$ and $|b^1\Sigma^+_g, N_a=0\rangle$ rotational states of O$_2$. (right orange box): Rotational states $|X^1\Sigma^+_g, N_b\rangle$ of the collision partner N$_2$. (c)~The angle-averaged potential energy surfaces (PESs) for the $X$ (green curve) and $b$ (red dashed curve) terms of O$_2$. (d)~Partial-wave contributions to the generalized cross-section $\text{Re}\,\sigma^{(q)}_0$ at high collisional kinetic energy for the N$_2$-perturbed target transition (black solid line). Also shown is the analogous result for the H$_2$-He case (gray dashed line).
    }
\end{figure}

\paragraph{Introduction}The oxygen (O$_2$) \textit{A}-band comprises rovibronic transitions connecting vibrational ground states of the $X^{3}\Sigma^{-}_{g}$ and $b^{1}\Sigma^{+}_{g}$ electronic terms of O$_2$ (Fig.~\ref{fig: fig 1}~(a)). It plays a vital role in studies of the Earth's atmosphere, where the high concentration of O$_2$ and long absorption path lengths allow optical saturation of the weak spin-forbidden \textit{A}-band lines~\cite{long_hodges2012}. This band spans an otherwise highly transparent near-infrared window around 760~nm, making it an important tool in atmospheric science with numerous calibration-free retrieval applications~\cite{yamamoto1961, barton1986, kuze1994, OCO-2, diedenhoven2005_SCIAMACHY,  corradini2006_SCIAMACHY, guanter2010, long_hodges2012, SCIAMACHY}. However, due to the prevalence of molecular collisions in the atmosphere, the accuracy of O$_2$ retrievals is limited by uncertainty in the available collisional lineshape parameters. Additionally, accurate extraction of atmospheric properties from O$_2$ spectroscopic measurements requires using beyond-Voigt lineshape profiles~\cite{ritter1987, ray1997_dicke, ray1997, schermaul1999, brown2000, tran2006, tran2008, robichaud2008, predoi-cross2008_O2N2, predoi-cross2008_O2O2,  long_hodges2012, drouin2017, mendonca2019, fleurbaey2021, adkins2025_survey, stevenson2025, klemm2025}.

Historically, collisional lineshape parameters for the O$_2$ \textit{A}-band have been investigated primarily through experiment~\cite{nakazawa1982, ritter1987, ray1997_dicke, ray1997, schermaul1999, pope2000, brown2000, yang2000, hill2003, tran2006, robichaud2008, predoi-cross2008_O2N2, predoi-cross2008_O2O2, drouin2017, adkins2025_survey, stevenson2025, klemm2025, merrell2026}, with only a few semi-empirical~\cite{tran2006, tran2008} or classical molecular dynamics calculations~\cite{hartmann2013,tran2019}. The only fully quantum-mechanical treatment was given by Grimminck \textit{et al.}~\cite{grimminck2014} for the case of pressure broadening and shift of the \textit{A}-band lines due to collisions with He($^1S_0$) atoms. To date, no theoretical studies have investigated collisional lineshape effects in the O$_2$ \textit{A}-band perturbed by major atmospheric constituents entirely from first principles of quantum mechanics, nor have they provided any rigorous \textit{ab initio} calculations of beyond-Voigt lineshape parameters for such systems.

In this Letter, we report a fully \textit{ab initio} calculation of lineshape parameters for the P1P1 {$^{{16}}{\rm{O}}_2$} \textit{A}-band transition perturbed by collisions with $^{14}{\rm{N}}_2$ (Fig.~\ref{fig: fig 1}~(b)). Henceforth, all computed and measured spectra correspond to this absorption line and molecular system. We present a treatment of a molecular electronic transition collisionally perturbed by another molecule from first principles which simultaneously accounts for the electronic structure, scattering dynamics and beyond-Voigt lineshape effects. To validate the theory, we perform accurate measurements of the spectra across a broad pressure range relevant to atmospheric conditions. We show that the \textit{ab initio} and experimental lineshapes agree to better than 0.6\%, although there are discrepancies in the individual parameters. We discuss how limitations of the electronic structure calculations of the O$_2$-N$_2$ interaction potential preclude accurate prediction of the collisional shift parameter. We also estimate \textit{ab initio} parameter bias by comparison to experiment, and we describe how higher-order \textit{ab initio} lineshape parameters can be used to constrain fits of experimental spectra.


\paragraph{Theoretical methodology}To investigate the N$_2$-induced collisional perturbations of the O$_2$ $A$-band, the O$_{2}$-N$_{2}$ quantum scattering problem was solved for the ground ($\eta = X$) and excited ($\eta = b$) electronic states of the absorbing O$_{2}$. To this end, two \textit{ab initio} potential energy surfaces (PESs) $V_{\eta}$ were calculated for $^{16}\text{O}_{2}(\eta)$-$^{14}\text{N}_{2}(X^{1}\Sigma^{+}_{g})$ interactions using explicitly correlated unrestricted coupled-cluster theory~\cite{knizia2009}. The present work reports a new $V_{b}$, while using the $V_{X}$ from Ref.~\cite{gancewski2021}. 

In principle, $V_{X}$ and the associated O$_{2}$-N$_{2}$ scattering dynamics depend on the total electronic spin of O$_{2}$. However, since the collisional kinetic energies relevant for atmospheric conditions ($\approx$ 100~cm$^{-1}$ to 300~cm$^{-1}$) are much larger than fine-structure corrections to rotational energies of O$_{2}$ ($\approx$ 2~cm$^{-1}$, see Fig.~\ref{fig: fig 1}~(b)), the spin dependence of the O$_{2}$-N$_{2}$ Hamiltonian may be neglected in quantum scattering calculations. The collisional problem was therefore formulated in terms of a spin-independent Hamiltonian (O$_{2}$ and N$_{2}$ labeled as $a$ and $b$, respectively)
\begin{equation}
    H_{\eta} = -\frac{\hbar^{2}}{2\mu R}\frac{\partial^{2}}{\partial R^{2}}R + \frac{\hbar^{2}\bs{l}^{2}}{2\mu R^{2}} + \mathcal{H}_{a(\eta)} + \mathcal{H}_{b} + V_{\eta}(R, \hat{\bs{r}}_{a}, \hat{\bs{r}}_{b}) \rm{,}
    \label{eq: Hamiltonian}
\end{equation}
where $R$ is the distance between nuclear mass centers of the colliding molecules, $\mu$ is the reduced mass, $\bs{l}$ is the relative orbital angular momentum and $\mathcal{H}_{i}$ is the rotational Hamiltonian of the $i$-th isolated monomer whose orientation relative to the axis taken along $R$ is given by a unit vector $\hat{\bs{r}}_{i}$. The dependence of $V_\eta$ on the monomer orientations leads to rotationally inelastic collisions. The angle-averaged PESs for both $\eta$ terms of O$_2$ are shown in Fig.~\ref{fig: fig 1}~(c).

Exact coupled-channels quantum scattering calculations were then performed for each $\eta$ using the \verb|BIGOS| code~\cite{jozwiak2024_AA, BIGOS_code}, and appropriate scattering boundary conditions were imposed~\cite{green1975} to obtain spin-independent $S$-matrix elements. The non-zero total electronic spin of O$_2$ was accounted for by transforming the $S$-matrices to the representation that includes spin explicitly~\cite{offer1994, gancewski2021} and is thus capable of correctly labeling the internal monomer states involved in the \textit{A}-band transitions (Fig.~\ref{fig: fig 1}~(b)).

\begin{figure*}
    \centering
    \includegraphics[width=\linewidth]{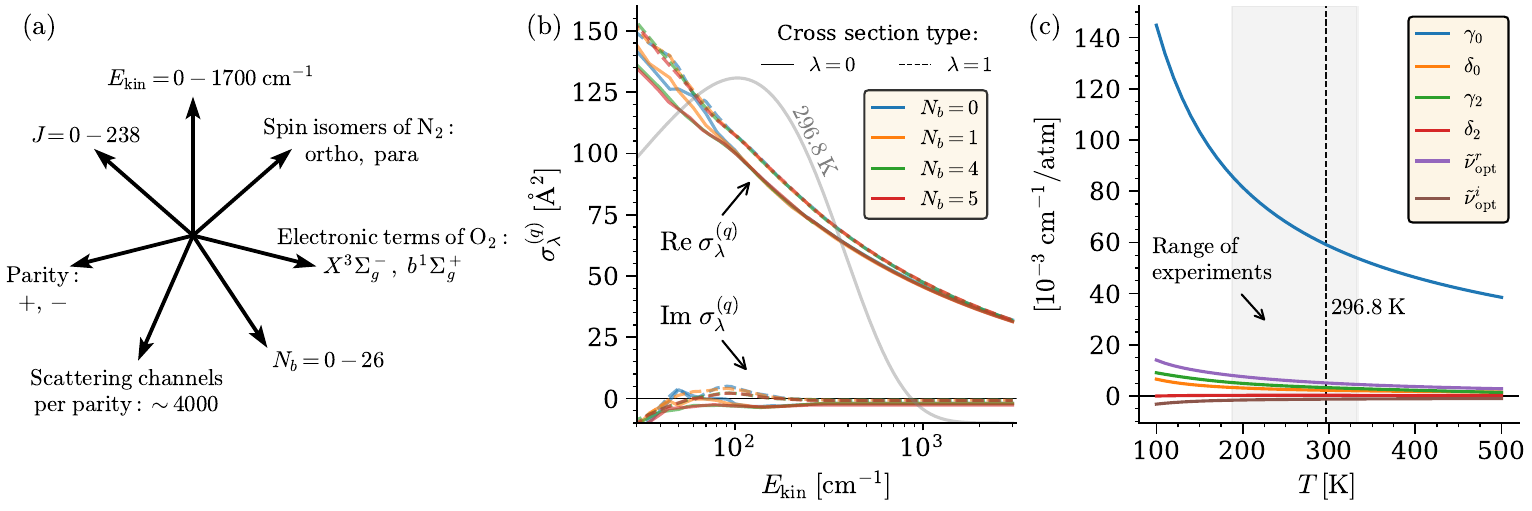}
    \caption{
        \label{fig: fig 2}
        (a)~Complexity of the quantum scattering problem. (b)~Generalized cross-sections $\sigma^{(q)}_{\lambda}$ as a function of collisional kinetic energy $E_\text{kin}$ and rotational state $N_b$ of the perturbing N$_2$ (only four examples of $N_b$ are shown, for clarity). The gray curve is the Maxwell distribution (in arbitrary units) at temperature $T = 296.8$~K. (c)~\textit{Ab initio} collisional lineshape parameters (Table~\ref{tab: tab 1}) as a function of temperature, with $T = 296.8$~K indicated by a vertical dashed line. The theoretical approach provides accurate lineshape parameters over a substantially wider temperature range than has been reported experimentally (gray vertical strip)~\cite{nakazawa1982, ritter1987, ray1997_dicke, ray1997, schermaul1999, pope2000, brown2000, yang2000, hill2003, tran2006, robichaud2008, predoi-cross2008_O2N2, predoi-cross2008_O2O2, drouin2017, adkins2025_survey, stevenson2025, klemm2025, merrell2026}.
    }
\end{figure*}

The high conceptual and computational complexity of the scattering problem~\cite{jozwiak_wcislo2026} presents a major challenge, illustrated schematically in Fig.~\ref{fig: fig 2}~(a), due to the accuracy requirements of atmospheric lineshape modeling. First, the large reduced mass of the O$_2$-N$_2$ complex and the relevant collisional kinetic energies require an expansive partial-wave basis substantially larger than for simpler molecule-atom cases like H$_2$-He (Fig.~\ref{fig: fig 1}~(d)). Second, the O$_2$-N$_2$ scattering dynamics are dictated by two highly anisotropic PESs coupling distant rotational states $N_a$ and $N_b$ of the monomers which, when combined with the small rotational constants of both molecules, demands an immense basis set of internal states (up to $50$ $\{N_a, N_b\}$ pairs). The partial-wave and internal state requirements result in a large number of scattering channels ($\approx$ 4000 per parity) that must be included in the \textit{ab initio} $S$-matrix calculations.

The transformed $S$-matrices were used to compute the complex-valued generalized cross-sections $\sigma^{(q)}_\lambda$~\cite{monchick1986, olejnik2023, jozwiak_wcislo2026} necessary to calculate the collisionally perturbed lineshape parameters. In contrast to the more familiar differential scattering cross-sections, the generalized cross-sections depend on the rank of a $2^q$-pole transition moment and on the index $\lambda = 0, 1$ associated with transport phenomena (e.g., Dicke narrowing), complicating the dependence on $S$-matrix elements. For the \textit{A}-band transitions driven by the O$_2$ magnetic dipole moment, $q = 1$. 

Figure~\ref{fig: fig 2}~(b) shows the \textit{ab initio} cross-sections $\sigma^{(q)}_\lambda$ for the N$_2$-perturbed P1P1 O$_2$ \textit{A}-band transition. It is clearly seen that the collisional effects are dominated by the real parts, $\text{Re}\,\sigma^{(q)}_\lambda$, which are nearly two orders of magnitude larger than the imaginary parts, $\text{Im}\, \sigma^{(q)}_\lambda$. The calculations cover more than 99.9$\%$ of the statistical distributions for the relative O$_2$-N$_2$ speeds and rotational states of the perturbing N$_2$ at atmospherically relevant temperatures ($\approx$ 300~K) in order to meet the requirements of atmospheric applications. This translates to a broad range of collisional kinetic energies, $E_\text{kin}$, and rotational states, $N_b$, with exceptionally high upper limits ($E_\text{kin} \approx$ 1700~cm$^{-1}$, $N_b \approx 26$), see Fig.~\ref{fig: fig 2}~(a). We note that it is possible to cover large values of $E_\text{kin}$ and $N_b$ using the extrapolated $\sigma^{(q)}_\lambda$ without loss of accuracy, see End Matter.

The cross-sections relate the microscopic intermolecular interactions to the macroscopic transport-relaxation phenomena in the absorbing gas, allowing the computation of collisional parameters of the modeled lineshape profiles. Here, we generate the \textit{ab initio} collisional lineshape parameters of the modified Hartmann-Tran (mHT) profile~\cite{wcislo2025} for the N$_2$-perturbed O$_2$ target line. Besides the standard Voigt collisional parameters (speed-averaged pressure broadening, $\Gamma_0$, and shift, $\Delta_0$), the mHT model accounts for the leading beyond-Voigt effects: speed dependence of broadening and shift ($\Gamma_2$ and $\Delta_2$, respectively), and velocity-changing collisions quantified by the complex-valued Dicke parameter, $\nu_\text{opt}$~\cite{wcislo2025}. The mHT lineshape parameters were determined by averaging $\sigma^{(q)}_\lambda(E_\text{kin}; N_b)$ over $E_\text{kin}$ and $N_b$. In particular, $\lambda = 0$ cross-sections were used to compute the speed-independent pressure broadening and shift~\cite{monchick1986, schaefer1992, wcislo2018}
\begin{equation}
    \Gamma_0 + \text{i}\Delta_0 = \frac{1}{2\pi c}\frac{p}{k_\text{B}T}\overline{v}_r \sum_{N_b} \rho_{N_b} \int\limits_{0}^{\infty} \text{d}{x}\, x e^{-x} \sigma^{(q)}_0(xk_\text{B}T; N_b) \rm{,}
    \label{eq: Gamma_0 + iDelta_0}
\end{equation}
and their leading speed dependence~\cite{wcislo2021}
\begin{align}
    &\Gamma_2 + \text{i}\Delta_2 = \frac{1}{2\pi c} \frac{p}{k_\text{B}T} \frac{v^*_b}{\sqrt{\pi}} e^{-y^2} \sum_{N_b} \rho_{N_b}\int\limits_{0}^{\infty}\text{d}x \notag
    \\
    & \left[2x\cosh{(2xy)} - \left(\frac{1}{y} + 2y\right)\sinh{(2xy)}\right] x^2 e^{-x^2} \notag
    \\
    &\sigma^{(q)}_0\left(\mu(xv_b^*)^2/2; N_b\right) \rm{,}
    \label{eq: Gamma_2 + iDelta_2}
\end{align}
where $p$ is the pressure, $T$ is the temperature, $\rho_{N_b}$ is the Boltzmann distribution of rotational states of the perturbing N$_2$, $\overline{v}_r$ is the average relative O$_2$-N$_2$ speed, $v_b^*$ is the most probable N$_2$ speed, and $y^2\equiv m_b/m_a$ is the perturber-to-absorber mass ratio. On the other hand, $\lambda = 1$ cross-sections were used for calculating the Dicke parameter~\cite{monchick1986, schaefer1992, wcislo2021}
\begin{align}
    &\nu_\text{opt} \equiv \nu^r_{\mathrm{opt}}+\text{i}\nu^i_{\mathrm{opt}}= \frac{1}{2\pi c} \frac{p}{k_\text{B}T} \frac{m_b}{m_a+m_b} \overline{v}_r \sum_{N_b} \rho_{N_b} \notag\\
    &\int\limits_{0}^{\infty}\text{d}{x}\, \left[\frac{2}{3}x\sigma^{(q)}_1(xk_\text{B}T; N_b) - \sigma^{(q)}_0(xk_\text{B}T; N_b)\right] x e^{-x}.
    \label{eq: Dicke}
\end{align}

\begin{figure*}
    \includegraphics[width=\linewidth]{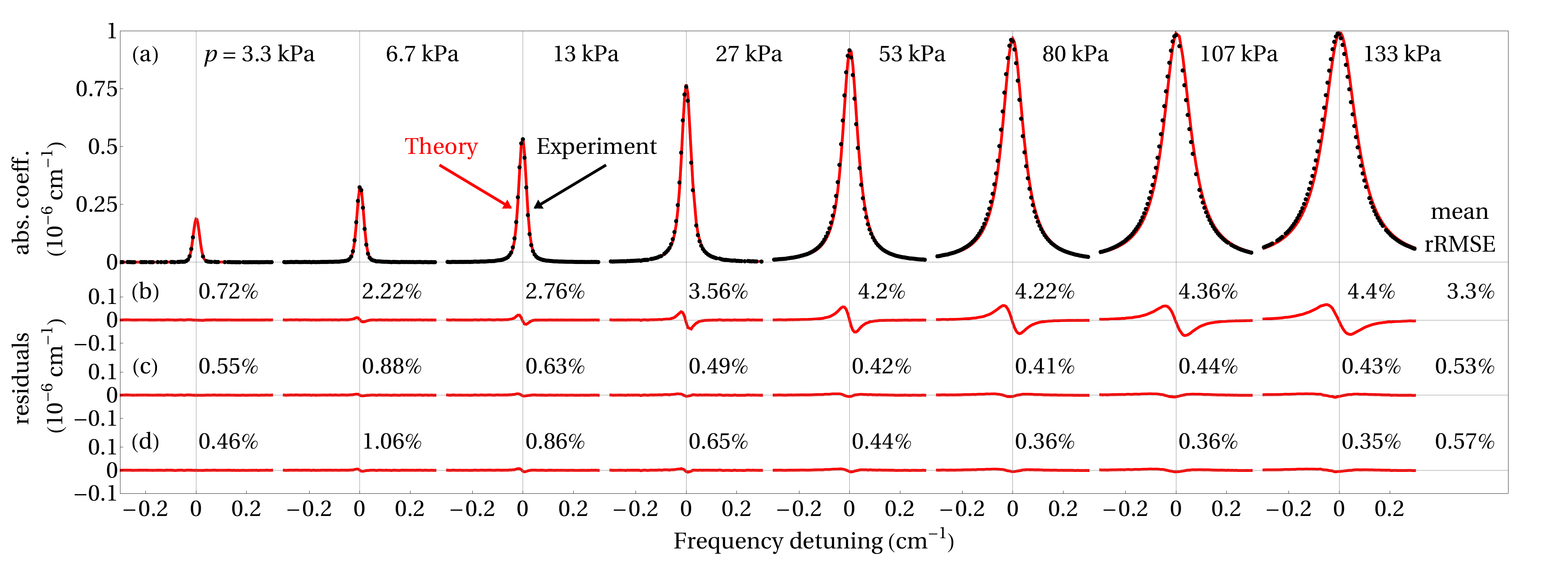}
    \caption{
        \label{fig: fig 3}
        Experimental validation of our quantum \textit{ab initio} calculations of the lineshapes. (a) Experimental (black points) and theoretical (red lines) shapes of the N$_2$-perturbed target line at $T=296.8$~K and eight pressures. (b) Differences between the experimental and fully \textit{ab initio} theoretical spectra. (c) The same as panel (b) but the collisional line shift parameter in theoretical spectra is adjusted to experimental spectra. (d) The same as panel (b) but in the \textit{ab initio} calculations the potential energy surface for the $b^{1}\Sigma^{+}_{g}$ state was multiplied by 1.0182. The relative root-mean-square error (rRMSE), shown above each curve in panels (b) - (d), was calculated within $\pm$ 1 full width at half maximum (FWHM) and normalized to the profile amplitude.  The pressure-averaged  rRMSE  are reported in the right side of the figure.
    }
\end{figure*}

Figure~\ref{fig: fig 2}~(c) shows the per-unit-pressure \textit{ab initio} lineshape parameters (denoted $\gamma_0$, $\delta_0$, $\gamma_2$, $\delta_2$, $\tilde{\nu}^r_\text{opt}$ and $\tilde{\nu}^i_\text{opt}$) as a function of temperature, where the values at the experimental temperature, 296.8~K, are reported in Table~\ref{tab: tab 1}~(a). The collisional effects are dominated by $\gamma_0$, an expected consequence of the two-orders-of-magnitude difference between the real and imaginary parts of $\sigma^{(q)}_0$ (cf. Fig.~\ref{fig: fig 2}~(b) and Eq.~\eqref{eq: Gamma_0 + iDelta_0}). While one might expect $\tilde{\nu}^r_\text{opt}$ to be similarly large, Eq.~\eqref{eq: Dicke} involves the difference between two cross-sections of comparable magnitude therefore yielding a substantially smaller value than $\gamma_0$. The calculated collisional parameters were used to generate the fully \textit{ab initio} mHT lineshape profiles, shown in Fig.~\ref{fig: fig 3}~(a) as red curves for eight pressures spanning from the low-pressure Doppler limit to the collision-dominated high-pressure regime.

\paragraph{Measurements}To benchmark the \textit{ab initio} calculations, high-precision measurements of the N$_2$-perturbed target transition were made using a frequency-stabilized cavity-ring down spectroscopy system, previously described in Refs.~\cite{stevenson2025, adkins2025}. This system provided enough sensitivity to enable use of a dilute oxygen sample ($0.25 \%$ of O$_2$ in N$_2$) and yielded a nearly distortion-free measurement of lineshapes ~\cite{fleurbaey2021}. These experimental attributes resulted in high fidelity spectra against which the \textit{ab initio} calculations were compared.

\begin{table}
    \caption{
        \label{tab: tab 1}
        Pressure-normalized collisional lineshape parameters for the N$_2$-perturbed target transition at $T=296.8$~K. All  reported quantities have units of $10^{-3}$~cm$^{-1}$/atm (1 ~cm$^{-1}$/atm = 0.295 872 15~MHz/Pa). (a) Fully \textit{ab initio} calculations. (b) \textit{Ab initio} calculations with excited-state $V_b$ PES multiplied by 1.0182.  (c) \textit{Ab initio} calculations (same as column (a)) with collisional shift $\delta_0$ adjusted to experimental spectra, denoted by (*). (d) Results from least-squares multi-spectrum fit of the mHT profile to measured spectra giving an rRMSE = 0.06\%. (e) Same as (d) with two constraints: $\gamma_0$ equal to the experimental value in column (d), and $\gamma_2/\gamma_0 = 0.057$ given by the ratio of corresponding theoretical values in column (b). Varying the remaining four fitted parameters gives an rRMSE =  0.08\%. Values in parentheses are the statistical uncertainties in the last reported digit. Constrained parameters do not report a statistical uncertainty.
    }
    \begin{ruledtabular}
    \begin{tabular}{lddddd}
        & \multicolumn{3}{c}{Theory} & \multicolumn{2}{c}{Experiment} \\
        & \multicolumn{1}{c}{(a)} & \multicolumn{1}{c}{(b)}
        & \multicolumn{1}{c}{(c)} & \multicolumn{1}{c}{(d)} & \multicolumn{1}{c}{(e)}\\
        \hline
        $\gamma_{0}$ & 59.1 & 59.3 & 59.1 & 60.33(2) & 60.33(-)\\
        $\delta_{0}$ &  2.2 & -4.0 & -4.0\,\rm{(*)} & -3.58(2) & -3.48(2)\\
        $\gamma_{2}$ &  3.3 &  3.4 &  3.3 & 4.35(8) & 3.46(-)\\
        $\delta_{2}$ &  0.3 & -0.8 &  0.3 & -0.69(8) & -0.34(11)\\
        $\tilde{\nu}^r_{\mathrm{opt}}$ &  5.1 &  5.0 & 5.1 & 2.1(2) & 5.54(12)\\
        $\tilde{\nu}^i_{\mathrm{opt}}$ & -1.2 & -1.5 & -1.2 & 0.7(3) & -0.5(4)\\
    \end{tabular}
    \end{ruledtabular}
\end{table}

\paragraph{Comparison of experiment and theory}The experimental N$_2$-perturbed lineshapes are shown in Fig.~\ref{fig: fig 3}~(a) as black points. The differences between the experimental and theoretical spectra are shown in Fig.~\ref{fig: fig 3}~(b). Also shown are the corresponding values of the relative root-mean-square error (rRMSE), calculated within $\pm$ 1 full width at half maximum (FWHM) and normalized to the profile amplitude, which quantify the magnitude of the differences between theory and experiment \cite{slowinski2020}. The pressure-averaged rRMSE is 3.3\%. However, the differences between the experimental and theoretical lineshapes systematically exhibit odd symmetry about line center and increase with pressure, which indicates that the dominant source of the discrepancy is the collisional shift parameter, $\delta_0$. To confirm the impact of pressure shift, we simultaneously fit the mHT profile to all experimental spectra adjusting the value of $\delta_0$ while fixing all other collisional lineshape parameters to the \textit{ab initio} theoretical values, see Table~\ref{tab: tab 1}~(c). In Figure~\ref{fig: fig 3}~(c), we show the corresponding residuals, which confirm the differences in Fig.~\ref{fig: fig 3}~(b) can mostly be attributed to inaccuracy in the calculated collisional line-shift parameter. Noting that the collision-induced line shift only displaces the line center as a function of pressure, a key result is that the present methodology can predict the collision-induced shapes of a molecular line from fully \textit{ab initio} quantum calculations at the sub-percent level (a mean rRMSE = 0.57\%, see Fig.~\ref{fig: fig 3}~(d)). 

To extend the analysis, we consider the origin of the pressure shift discrepancy. The value of $\delta_0$ comes from the imaginary part of the generalized spectroscopic cross-section, see Eq.~\eqref{eq: Gamma_0 + iDelta_0}. We note that  this component is related to the difference between scattering phase shifts~\cite{jozwiak2018}, which may change sign~\cite{stankiewicz2020} and which depends on the number of closed channels included in the basis~\cite{jozwiak2021}. Consequently, $\text{Im}\, \sigma^{(q)}_0$ converges more slowly in numerical quantum scattering calculations than $\text{Re}\, \sigma^{(q)}_0$. The pressure shift is also much more sensitive to the details of the PES, which we observe in the quantum-mechanical case and which was realized already in the context of semi-classical scattering calculations~\cite{lamouroux2012_pt2, lamouroux2012_pt3}.  To illustrate the sensitivity of $\delta_0$ to the PES, we modified the PES of the excited state by 1.82\% (i.e., we scaled $V_b$ by a factor of 1.0182) and repeated the \textit{ab initio} calculations. The results are shown in Table~\ref{tab: tab 1}~(b) and in Fig.~\ref{fig: fig 3}~(d). The small modification of the PES considerably changed the calculated $\delta_0$, yielding a result close to the experimental value, while introducing negligible changes to the other lineshape parameters. An exception is the $\delta_2$ parameter whose value is directly correlated with $\delta_0$, but for which we found had a negligible influence on the lineshape. 

\paragraph{Parameter correlation} A long-standing challenge in spectroscopic reference data is the physical interpretation of beyond-Voigt parameters obtained from fits of parameterized profiles to measured spectra~\cite{long_hodges2012, cygan2012, cygan2013, wcislo2016, adkins2022, fleurbaey2021, wcislo2025}. When fitting beyond-Voigt profiles to experimental spectra, substantial numerical correlation between parameters occurs because of their similar influence on the profile ~\cite{adkins2022}. In practice, various combinations of correlated parameters yield statistically indistinguishable profiles when residuals are smaller than the background noise. Therefore, unconstrained analyses typically lead to unphysical or biased fitted parameters. Typically, parameter correlation is reduced by employing a multi-spectrum fitting approach over a range of pressure and temperature ~\cite{benner1995, pine2001}. Nevertheless, the ability to completely disentangle parameter correlation is limited by the finite signal-to-noise ratio and the lack of more general models that constrain the relative magnitudes of the relevant beyond-Voigt lineshape terms ~\cite{fleurbaey2021}. The determination of fitted mHT parameters is further complicated by other higher-order effects, such as line mixing and intensity depletion~\cite{payne2020, adkins2025_survey, tran2026}, that introduce additional model parameters.

\paragraph{Estimating bias in calculated lineshape parameters and implementing advanced constraints} Our measurements of the speed-independent pressure broadening parameter have combined relative uncertainties of less than 0.1\%, as shown in Table~\ref{tab: tab 1}~(d). This result is a consequence of the high signal-to-noise ratio of the measurements and the fact that this fit-derived quantity is weakly correlated with the other higher-order mHT parameters. Therefore, we expect the difference between the calculated and measured speed-independent and speed-dependent collisional broadening parameters to be dominated by bias in our calculation of Re $\sigma_0^{(q)}$ (Eqs.~\eqref{eq: Gamma_0 + iDelta_0}--\eqref{eq: Gamma_2 + iDelta_2}). Comparing the $\gamma_0$ values in columns (b) and (d) of Table~\ref{tab: tab 1}, we find that the ratio of measured-to-calculated values is 1.017. If we attribute this discrepancy to systematic error in Re $\sigma_0^{(q)}$, then the calculated value of $\gamma_2$ would also be underestimated by 1.7\%.

With $\gamma_0$ and $\gamma_2$ constrained by comparison to experiment and theory (Table~\ref{tab: tab 1}~(e)), the remaining mHT parameters were obtained from multi-spectrum fits to the measured spectra. This analysis yielded an average rRMSE value of 0.08\%, which is comparable to that obtained with the unconstrained fit results (0.06\%), shown in Table~\ref{tab: tab 1}~(d). The resulting $\tilde{\nu}^r_{\mathrm{opt}}$ value is 10.8\% greater than the theoretical value. Recall that $\tilde{\nu}^r_{\mathrm{opt}}$ depends on the difference between two cross-section terms of similar magnitude, see Eq.~\eqref{eq: Dicke},  making its prediction more uncertain than that of $\gamma_2$. As expected, this constrained mHT fit has a much larger impact on $\delta_2$ and  $\tilde{\nu}^i_{\mathrm{opt}}$, yielding a negative value for the latter term consistent with the theoretical prediction, albeit with a relatively large uncertainty. To the best of our knowledge, this represents the most reliable determination of collisional lineshape parameters for O$_2$ to date, achieving exceptionally low fit residuals while preserving the physical consistency of the retrieved parameters.

In addition to providing accurate $\textit{ab initio}$ lineshape parameters in the absence of experimental results, the present theory provides additional constraints for fitting the mHT profile to experimental spectra. Here, this methodology has been demonstrated for the P1P1 transition, but it can be applied to other \textit{A}-band lines, other O$_2$ bands and other diatomic molecules perturbed by O$_2$ and N$_2$. In future experimental analyses, by constraining $\gamma_2$ as discussed above, its numerical correlation with the other mHT parameters, especially $\tilde{\nu}^r_{\mathrm{opt}}$, can be substantially reduced.  This approach, which leverages new theoretical constraints, will yield more physically consistent mHT parameter retrievals.

\paragraph{Conclusions}The present study demonstrates an exceptional level of agreement between \textit{ab initio} and measured collision-induced shapes of a rovibronic molecular transition. The strength of the first-principles approach is seen in Fig.~\ref{fig: fig 2}~(c). The calculations not only provide the beyond-Voigt lineshape parameters, but also cover a wide range of temperatures which are difficult to experimentally realize. The future of accurate molecular lineshape data will be heavily based on the interplay between \textit{ab initio} quantum calculations and accurate spectroscopic measurements. The calculations can provide accurate temperature-dependent lineshape parameters originating from the real parts of the spectroscopic cross-sections. Experiments can be used to refine the calculations of the complex-valued cross-sections to more accurately account for the relative magnitudes of the lineshape terms.  By coupling \textit{ab initio} calculations with high-resolution measurement benchmarking, it will be possible to generate comprehensive and reference-grade spectroscopic databases, which are necessary for atmospheric studies of Earth and other planetary systems ~\cite{HITRAN2024}. Beyond the context of atmospheric studies, the present methodology has basic-science applications as it provides a sensitive tool for evaluating \textit{ab initio} PESs by probing the imaginary part of the cross-sections shown in this work, challenging the present state-of-the-art electronic structure calculations~\cite{slowinski2020, slowinski2022, jozwiak2024}.

\begin{acknowledgments}
\paragraph{Acknowledgments}The authors would like to thank Daniel Barker (NIST) and Eisen Gross (NIST) for commenting on the manuscript. The research was supported by the National Science Centre, Poland, Project Nos.: 2021/43/O/ST2/00211 (MG) and 2022/46/E/ST2/00282 (HJ, NS, PW). RD and EQS. are supported by the United States Department of Energy (DOE), Grant No. DE-SC0025420. We gratefully acknowledge Polish high-performance computing infrastructure PLGrid (HPC Center: ACK Cyfronet AGH) for providing computer facilities and support within computational grant no. PLG/2026/019877. Created using resources provided by Wroclaw Centre for Networking and Supercomputing (http://wcss.pl). The research is part of the program of the National Laboratory FAMO in Toruń, Poland.
\end{acknowledgments}

\paragraph{Data availability}Experimental spectra are available at \url{https://data.nist.gov/}. Additional data are available
from the authors upon reasonable request.

\bibliography{bibliography}

\section*{End Matter}

\paragraph{Additional calculation details}Our calculations revealed that the dominating real part of the spectroscopic cross-sections, $\text{Re}\, \sigma^{(q)}_\lambda$, follows a power law in $E_\text{kin}$ for large kinetic energies. We therefore performed the \textit{ab initio} calculations up to $E_\text{kin} = 300$~cm$^{-1}$ and used the high-energy ``tail'' of the calculated $\text{Re}\, \sigma^{(q)}_\lambda$ to fit a power-law dependence and extrapolate $\text{Re}\, \sigma^{(q)}_\lambda$ to higher kinetic energies. We observed no such dependence on $E_\text{kin}$ for the imaginary part, $\text{Im}\, \sigma^{(q)}_\lambda$, two orders of magnitude smaller than the real part. Instead, the calculated $\text{Im}\, \sigma^{(q)}_\lambda$ stabilize at high $E_\text{kin}$, approaching small constant values near zero. Due to lack of a more definite functional dependence on $E_\text{kin}$, we extrapolated $\text{Im}\, \sigma^{(q)}_\lambda$ to higher kinetic energies using the stabilized constant values. We verified that due to the two-orders-of-magnitude difference between the real and imaginary parts, the extrapolation scheme does not influence the accuracy of the \textit{ab initio} lineshape; the primary factor is the quality of the O$_2$-N$_2$ interaction PES, as demonstrated in Table~\ref{tab: tab 1} and Fig.~\ref{fig: fig 3}.

Furthermore, for all calculated $\sigma^{(q)}_\lambda$, we observed that the cross-sections become almost independent of the rotational state of the perturbing N$_2$ already for $N_b \gtrsim 4$ (Fig.~\ref{fig: fig 2}~(b)). Consequently, we performed the \textit{ab initio} calculations up to $N_b = 9$ and then---depending on the spin isomer of N$_2$ (ortho vs. para)---we extrapolated $\sigma^{(q)}_\lambda$ to higher rotational quantum numbers $N_b$ using the \textit{ab initio} cross-sections calculated for $N_b = 8, 9$.

Finally, we emphasize that, when calculating the lineshape parameters (Eqs.~\eqref{eq: Gamma_0 + iDelta_0}--\eqref{eq: Dicke}), we use all \textit{ab initio} values of $\sigma^{(q)}_\lambda$ and switch to the extrapolated cross-sections only when the \textit{ab initio} values reach their $E_\text{kin}$ and $N_b$ limits.

\paragraph{Additional measurement details}This work reports high-precision measurements of the N$_2$-perturbed target transition made using a frequency-stabilized cavity-ring down spectroscopy system, previously described in Refs.~\cite{stevenson2025, adkins2025}. The optical cavity, HeNe length-stabilization servo, frequency agile rapid scanning, and temperature measurement scheme remain the same as those detailed in Ref.~\cite{adkins2025}. The current measurements included optical frequency comb (OFC) referencing of the frequency axis and flowed samples, which differ from the previous descriptions Refs.~\cite{stevenson2025, adkins2025}. 

The current measurements referenced the cavity mode axis at each pressure to the visible output of a self-referenced, femtosecond optical frequency comb (OFC) (referenced to a global positioning satellite disciplined Rubidium oscillator) using a heterodyne beat signal.  A free-space optical band pass filter and RF band pass filters were used to ensure the laser carrier frequency was only beat with one comb tooth.  Heterodyne beat note measurements were made at discrete cavity resonant frequencies across the O$_2$ \textit{A}-band, providing absolute frequency anchors to the broadband cavity mode axis, with an absolute frequency uncertainty less than 300~kHz.  

The 0.25\% of O$_2$ in N$_2$ sample measured in this work was under flow where the intra-cavity pressure was stabilized ($\approx$ 1~Pa over broadband spectrum) using a proportional servo operating on a voltage-controlled valve placed between the output of the optical cavity and a vacuum pump.  The volumetric flow-rate was between 10 and 70~std.~cm$^3$~min$^{-1}$.  The pressure was measured using a resonant silicon manometer with a full-scale range of 130~kPa, calibrated against a NIST primary pressure transfer standard, and corrected to account for flow-induced pressure drop between the manometer and the sample volume, where the maximum magnitude of the correction was less than 20~Pa.  

\end{document}